# Observation of high-$T_c$ superconductivity in inhomogeneous combinatorial ceramics

Mitra Iranmanesh[a], Nikolai D. Zhigadlo[a,*], Thanaporn Tohsophon[a], John R. Kirtley[b], Wilfried Assenmacher[c], Werner Mader[c], Jürg Hulliger[a,**]

[a]Department of Chemistry and Biochemistry, University of Bern, CH-3012 Bern, Switzerland

[b]Geballe Laboratory for Advanced Materials, Stanford University, Palo Alto, California 94305, U.S.A.

[c]Institute of Inorganic Chemistry, University of Bonn, D-53117 Bonn, Germany

**Abstract**

A single-sample synthesis concept based on multi-element ceramic samples can produce a variety of local products. When applied to cuprate superconductors (SC), statistical modelling predicts the occurrence of possible compounds in a concentration range of ∼50 ppm. In samples with such low concentrations, determining which compositions are superconducting is a challenging task and requires local probes or separation techniques. Here, we report results from samples with seven components: $BaO_2$, $CaCO_3$, $SrCO_3$, $La_2O_3$, $PbCO_3$, $ZrO_2$ and $CuO$ oxides and carbonates, starting from different grain sizes. The reacted ceramics show different phases, particular grain growth, as well as variations in homogeneity and superconducting properties. High-$T_c$ superconductivity up to 118 K was found. Powder x-ray diffraction (XRD) in combination with energy-dispersive spectroscopy (EDS), scanning transmission electron microscopy (STEM) can assign "Pb1223" and "$(Sr,Ca,Ba)_{0.7-1.0}CuO_2$" phases in inhomogeneous samples milled with 10 mm ball sizes. Rather uniform samples featuring strong grain growth were obtained with 3 mm ball sizes, resulting in $T_c$ ~70 K superconductivity of the "$La(Ba,Ca)_2Cu_3O_x$" based phase. Scanning SQUID microscopy (SSM) establishes locally formed superconducting areas at a level of a few microns in inhomogeneous superconducting particles captured by a magnetic separation technique. The present results demonstrate a new synthetic approach for attaining high-$T_c$ superconductivity in compounds without Bi, Tl, Hg, or the need for high-pressure synthesis.

# 1. Introduction

The synthesis of new high-$T_c$ superconductors is one of the most challenging tasks in solid-state chemistry [1]. Consequently, to produce novel unexpected phases, new synthetic procedures are continuously being attempted. Despite a number of approaches, including 2D combinatorial methods [2], kinetically controlled synthesis [3], strained artificial layers [4], etc., no real progress towards the goal of a transition temperature above 138 K has been reported. On the other hand, a steady technical demand for materials featuring a $T_c$ of 150 K (or higher) is still present, e.g., for the power-grid technology [5].

In this situation, one can conceive alternative procedures able to explore with a high probability all the possible reaction pathways. To this aim, we recently proposed [6-8] and developed [9,10] a new procedure, in which we start with a random mixture of a large number of solid materials to achieve a final single phase.

Within such a system of $N$ components local reactions can occur, depending on the availability or mobility of some starting materials $n$ ($n < N$). For small grain sizes (in the range of microns or even below) the local amount of each component $i$ ($i \in n$) will determine the stoichiometry of a possible compound. In the case in which large grain sizes (50 – 100 µm or larger) are provided, neck formation [11] may dominate, thus allowing all phases given by $n$ components at temperature T and pressure $p$ [12] to form.

Once phase formation has occurred, we will end up with samples of strongly intergrown crystallites, most of them non-superconducting. However, numerical simulations [2,8] have shown that such a combinatorial system should be able to produce any possible (T, $p$) phase in the range of at least 50 ppm. Therefore, further processing will either need (i) magnetic phase separation [13] or (ii) local phase and property analysis. Magnetic separation (i) is at present able to catch superconducting grains of size up to 20-50 µm at a level of ppm concentration [9-13]. Local probe analysis (ii) can rely on Lorentz microscopy [14] to demonstrate superconductivity for µm to nm sized crystallites, along with elemental or even compositional analyses. A third method (iii), which we will apply here, is scanning SQUID microscopy (SSM) [15]. SSM allows a spatial resolution down to a few hundred nm, without, however, gaining information on composition.

In the search for new cuprate materials of practical interest that do not contain Tl, Hg or Bi, high pressure syntheses starting from Ca, Sr, Ba and Cu have yielded $T_c$'s up to 110-115 K [16,17].

In the present work we rely mainly on these elements (Ca, Sr, Ba and Cu) plus minor additions of La, Zr and Pb. These last three elements are found to be necessary components in the combinatorial system in order to produce unexpected phases at normal pressure, so far only obtained by high pressure syntheses [18].

As mentioned above, we vary the grain size of starting materials in order to allow the chemical system to proceed along mechanisms (i) or (ii). For analysis we apply magnetic grain separation, scanning SQUID microscopy, bulk SQUID, X-ray and EDS measurements to characterize the superconducting phases.

## 2. Experimental details

Initial compositions were prepared to react the combinations of seven elements in different relative molar ratios and the best results were obtained for the following volume fractions $2BaO_2$, $2CaCO_3$, $2SrCO_3$, $0.25La_2O_3$, $0.1PbCO_3$, $0.5ZrO_2$ and $4.5CuO$. Mixtures of an attempted grain size were obtained by ball milling (FRITSCH, Planetary mono mill pulverisette 6) for one hour at 450 rpm in the presence of isooctane to provide a slurry. For milling we used either an (i) agate container with balls of 10 mm or (ii) a zirconia container with balls of 3 mm in diameter.

After drying, powders from (i), (ii) were pressed into disk shaped pellets ($\varnothing = 10$ mm, $d = 1.5$ mm) at a pressure of 900 MPa. Sample pellets were placed in an alumina ($Al_2O_3$) crucible and heat-treated within a horizontal quartz glass tube. Oxygen gas was flowed continuously at $p(O_2)$ of about 1 bar during the annealing process. Typically the temperature was first raised to 930 °C (just below the formation of liquids) within ∼ 6 h, then decreased to 800 °C with a cooling rate of 10 °C/h. Finally, the samples were cooled to room temperature at a rate of 100 °C/h. Although eutectic phases can appear in combinations of these components, surprisingly, liquid phase formation at T below ∼ 930 °C was not observed by DSC or optical microscopy.

The phase analysis of sintered samples was carried out by means X-ray powder diffraction using a STOE StadiP diffractometer and Cu $K_\alpha$ radiation. The morphology of crystallites and the elemental distribution within the bulk material were investigated by the Hitachi S-3000 N scanning electron microscope (SEM) equipped with a Noran SIX NSS 200 dispersive X-ray detector.

Temperature dependent dc magnetic susceptibility measurements of bulk samples were performed by a Magnetic Property Measurement System (MPMS-XL, Quantum Design) equipped with a reciprocating-sample option. The critical temperature ($T_c$) was evaluated as a linear extrapolation of the steepest part of the M(T) curves to M = 0.

For magnetic separation [11,13], as obtained samples were ground and sieved ($d \leq$ 300 µm). By this method we could capture superconducting particles of a minimum size of 20-50 µm at a temperature T ≥ 80 K. For this, particles were suspended in a liquid mixture of alkene gases. The procedure (for details see ref. [11,13]) allowed the removal of superconductive particles from the suspension for single grain analyses.

Scanning SQUID microscopy (SSM) imaging of the local magnetic moment and magnetic susceptibility was carried out on SC grains isolated by magnetic separation to detect local areas of superconductivity [15]. SSM measurements have been performed in two ways: (i) First from the as obtained surface and (ii) second after removing about 50 µm of a large grain by polishing. For these measurements a SQUID susceptometer produced in a joint collaboration between IBM and Stanford University was used [15]. It uses (Fig. 1) a pickup loop and a concentric, single-turn field coil that allows the application of a local field. Applying a low-frequency ac current to the coil and detecting the resultant flux through a pickup loop enabled the measurement of the mutual inductance between the pickup loop and field coil, which is modified by the presence of any magnetic sample. In the case of superconductivity, the repulsion of the applied field due to the Meissner effect results in a reduced flux near the sample and a negative total mutual inductance. It is estimated that the pickup loop area is about $1 \times 3$ µm$^2$. The SQUID was operated in a flux-locked feedback loop.

A second set of SSM measurements were conducted with a susceptometer providing a larger pickup loop for enhanced sensitivity. All susceptibility data was taken with 1 mA through the field coil at 1117 Hz, with current running from the "back" field coil / pickup loop pair to the "front" pair.

To gain elemental information regarding the superconducting particles isolated by the magnetic separation method, scanning transmission electron microscopy (STEM - Philips/FEI CM300UT-FEG) was used. For this the sample was prepared in a glove box and Ar atmosphere. It was washed with acetone to remove the vacuum grease from X-ray measurements and crushed in a mortar using hexane as a dispersing agent. A droplet of the dispersion was placed on the sample holder, a permeable carbon film, which was supported by an aluminium grid. By this technique we were able to investigate the possible phases in captured superconducting particles.

### 3. Results and discussion

Figure 2 presents the temperature dependencies of the bulk magnetic susceptibility, measured by SQUID. The observed curve (a) revealed a superconducting transition temperature of about 70 K for a sample prepared by 3 mm balls (labelled B-3). Many attempts were made to increase the $T_c$ by changing the molar ratios of starting components and by varying the temperature as well as time of annealing. Regardless of all these efforts, the $T_c$ of ~70 K was the prominent result achieved for samples B-3. When using balls of 10 mm this elemental composition lead to superconductivity up to 118 K (curve (b) in Fig. 2, labelled B-10).

X-ray diffraction patterns of sintered samples prepared by different ball sizes are presented in Fig. 3: The B-3 sample features sharp diffraction peaks. With the help of the WinXPoW [19] program the following phases could be indexed and identified: $Ca_3Sr_{11}Cu_{24}O_{41}$ (no. 00-048-1499, [20]), $La_2SrCu_2O_6$ (no. 00-043-0403, [21]), $SrZrO_3$ [22] and the superconducting phase of $LaBaCaCu_3O_x$ (no. 00-082-157, [23]) with $T_c(max)$ of 80 K [24]. In addition to these, a $Sr_{0.7}Ca_{0.3}CuO_x$ phase [25] featuring a high $T_c$ of 110 K [26] is apparently observed in the B-10 sample. The identification of other possible high-$T_c$ phases (e.g. Pb-1223) in sample B-10 was challenging due to a strong overlapping of numerous Bragg peaks in the range of 2 theta = 25 - 35 degrees (Fig. 3b). By varying the sintering temperature in the range of 800 - 930 °C (for 13 h), the $SrZrO_3$ phase preferably formed [22]. These findings allow us to conclude that the size of the particles in the starting mixtures substantially affects the phase formation in these multi-component systems. A smaller average grain size (dozens or hundreds of nanometers) may favour local products using up grains, whereas a larger size may give rise to products appearing also in necks. While both samples represent multiphase objects, high-$T_c$ superconductivity appears only in those prepared by mixtures milled with 10 mm balls. Given a diversity of interfaces, spontaneous nucleation may produce metastable and stable phases belonging to phase diagrams constituted by the number of components participating in locally occurring phase formation.

This is made visible by SEM: The surface morphology of sintered samples shown in Fig. 4 reveals that the B-10 sample (Fig. 4b and Fig. 4d) consists of aggregated particles in the 100 - 300 nm size range, whereas the B-3 sample (Fig. 4a and Fig. 4c) is strongly intergrown.

The elemental analyses of sintered B-3 and B-10 samples, performed by EDS in SEM mode, are shown in Table 1. EDS with $100 \times 100$ μm$^2$ area size ($\times 500$) was carried out at 4 different places. Standard deviation (SD) is used to determine the homogeneity of the B-3 and B-10 samples. Overall, the mixtures milled with a 10 mm ball size result in less homogeneous specimens than those milled with a 3 mm ball size.

We have attempted to reach the 118 K phase also by using a 3 mm ball size with various treatments, different Pb content, or sintering temperature. However, we obtained only $T_c \sim 70$ K, implying that the $LaBaCaCu_3O_x$ is the preferred phase under these conditions. In addition we could identify the $Sr_{0.7}Ca_{0.3}CuO_2$ phase (Fig. 2b) which features a $T_c \sim 110$ K (B-10).

Scanning squid microscopy (SSM) measurements were performed on a single $300 \times 200$ μm$^2$ superconducting B-10 particle, isolated by magnetic separation. An SSM overview image from the surface of an entire grain taken at 4.2 K is shown in Figure 5(a). As expected, superconducting inclusions are distributed irregularly throughout the sample and thus diamagnetic shielding was highly inhomogeneous. In the false color representation used in Figure 5, areas with stronger diamagnetic shielding are coded blue, while those with weaker diamagnetic shielding are coded red (see colorbars). A region indicated by the rectangular box in Figure 5(a), which included the highest diamagnetic signal, was selected for a study of the temperature dependence.

Susceptibility images of selected regions at variable temperature are displayed in Figure 5(b). It was challenging to measure the magnitude of the susceptibility at a particular sample point and temperature. This was due to a combination of several factors: First, we used a relatively small area pickup loop for these measurements to get good spatial resolution. This means that errors in the spacing between the sample and the sensor produced large variations in the diamagnetic signal observed. Second, the sample was quite rough. Some variation is apparent, for example, in the non-monotonic nature of the temperature dependent images in Figure 5(b). The region with the strongest shielding (the blue spot labelled 1) appears to be larger than the minimum spatial resolution of the microscope: Its susceptibility signal shows a full-width at half maximum 4 microns in length and 2 microns in width larger than the region labelled 2. The regions with the strongest diamagnetic shielding have the weakest temperature dependence up to about 105 K. Above this temperature the shielding drops quickly. On the other hand, the regions with relatively weak shielding show a more gradual falloff in susceptibility with temperature. Increasing temperature from 100 K to the critical point results in a steep susceptibility drop. Note that at 107 - 108 K, the point 2 disappears while the point 1 remains (see Figure 5(c)).

Further SSM measurements have been performed on the same 300 μm grain as in Figure 5 using a 5 μm inside radius loop susceptometer: Figure 6(a) shows a mosaic image of the diamagnetic susceptibility of the full sample at 4.2 K. In Figure 6(b), the region with high diamagnetic signal was selected to study its behaviour as a function of temperature. A plot of the maximum (blue) and minimum (red) diamagnetic shielding in the images of Figure 6(b) as a function of temperature is shown in Figure 6(c). It is interesting to note that the area with the strongest susceptibility at low temperature becomes weaker more rapidly than the area just above it, which represents the most strongly diamagnetic part close to $T_c$. As may be seen, the scanning SQUID susceptibility becomes comparable to the noise at about 115 K, compared with $T_c$ = 118 K evaluated from bulk SQUID magnetization data as described above. Figure 6(d) highlights the spots where 4 individual areas were picked out to plot the local susceptibility as a function of temperature (see Figure 6(e)).

From these results we can conclude that in these combinatorial samples, there are *local reaction centers*, producing the superconducting compounds observed here down to 2 - 4 μm in size. We note that areas corresponding to superconducting regions are not always uniform but often consist of even smaller and connected spots. This reflects a non-uniform superconductivity inside superconducting regions. One might imagine that during the reaction a superconducting area was initially formed at several reaction centers around some particular grains and their subsequent growth resulted in the formation of somewhat weaker superconducting links between them.

For further analysis to determine the high-$T_c$ phase in B-10, subsequent elemental compositions have been explored by EDX with a 5 nm beam size in conjunction with STEM on several captured particles, extracted by the magnetic separation technique. Lateral resolution and penetration depth of EDX strongly depend on beam-broadening and particle thickness. The thinner the sample and the smaller the beam-broadening, the more accurate the analytical information will be. Data on two of the typical particles and its relative elemental analysis performed at randomly chosen areas as marked in Figure 7 (a,b) are collected in Table 2.

According to elemental compositions and atomic percentages of different spots in Table 2, we conclude that the dominant superconducting compound basically stems from Sr, Ba, Ca, and Cu elements. For minor or trace level elements (La, Pb) errors arise due to the spectral interferences (some peaks may be hidden) and the form of spectral lines.

Despite all these difficulties we are able to identify phases for at least six spots with a certain approximation, i.e. #3, #4, and #6 in Fig. 7(a) and #1, #2, #3 in Fig. 7 (b) which may be assigned to the following phases "$(Sr,Ca,Ba)_{0.7-1.0}CuO_2$", "$(Pb,Sr,Cu)Sr_2(Ca,Sr)_2Cu_3O_x$"(Pb1223), and "$La(Ba,Ca)_2Cu_3O_x$". The first two of them indeed show a $T_c$ of 100 K when synthesized under *high pressure* conditions. We suppose that the $T_c \sim 118$ K in B-10 samples originates from the "Pb1223" phase. Note that a high atomic percentage of some elements, i.e. Ba in #3 and Zr, Cu in #4, probably stems from an electron penetration depth of several nanometers, resulting in atomic information from many atomic layers. Based on our X-ray and EDS analyses we can conclude that these phases can be made through a combinatorial approach, even though in small quantities.

## 4. Conclusions

In this paper we explored the phase space of samples made up of a mixture of seven (Ca, Sr, Ba, La, Zr, Pb, and Cu) components processed with different milling conditions. The size of grinding balls, number of balls and feed materials are the key to obtain starting mixtures of powders. 10 mm balls induce a low level of interaction between balls and feed components, resulting in a more inhomogeneous sample, while the small balls provide rather uniform samples with a higher grain connectivity produced by the heat treatment.

By XRD analysis on bulk samples we can identify the non-superconducting $Ca_3Sr_{11}Cu_{24}O_{41}$, $La_2SrCu_2O_6$ and $SrZrO_3$ phases and the low $T_c$ $La(Ba,Ca)_2Cu_3O_x$ superconducting phase in both samples. However, a $T_c$ up to 118 K, which so far was only obtained under high pressure conditions, is achieved particularly in an inhomogeneous sample processed by 10 mm ball sizes. Following EDS in combination with STEM we describe these phases ($T_c$ > 100 K) as "Pb-1223" and "$(Sr,Ca,Ba)_{0.7-1.0}CuO_2$". Scanning SQUID measurements revealed an inhomogeneous distribution of superconductivity produced by local reaction centers.

The present study suggests that there is plenty of room to vary the reaction conditions in multi-component systems to provide as many instances as possible, which could produce new compounds not accessible through conventional reaction systems. A key analytical technique for future work would be Lorentz microscopy to identify superconductivity as well as elemental information for even smaller areas.

**Acknowledgements**

The SQUID microscope used in this work was developed with the support of an NSF IMR-MIP contract, award number 0957616. J. K. was supported under the auspices of an NSF NSEC: Center for Probing the Nanoscale, award number 0425897. Part of this work was performed at the Stanford Nano Shared Facilities (SNSF), supported by the National Science Foundation under award ECCS-1542152. We thank M. Stir for collaboration on the early stage of this study. We also thank K. Krämer, B. Trusch and T. Shiroka for the use of the SQUID equipment, technical assistance, and helpful comments.

**Table 1.** Quantitative content of elements in atomic % and their standard deviation (SD) analysed by EDS in SEM mode of the samples prepared by 3 mm ball (B-3) and 10 mm ball size(B-10). The mean values of elements normalized to 4.5 Cu atoms lead to the following compositions: $Pb_{0.2}La_{0.4}Zr_{0.6}Ba_{1.8}Ca_2Sr_{3.1}Cu_{4.5}O_x$ (B-3), $Pb_{0.2}La_{0.4}Zr_{0.8}Ba_{2.3}Sr_{3.5}Cu_{4.5}O_x$ (B-10).

| Elements | B-3 Atomic (%) | | | | | | B-10 Atomic (%) | | | | | |
|---|---|---|---|---|---|---|---|---|---|---|---|---|
| | #1 | #2 | #3 | #4 | $\bar{x}$ | SD | #1 | #2 | #3 | #4 | $\bar{x}$ | SD |
| Ca-*K* | 17.0±0.5 | 15.5±0.5 | 16.4±0.5 | 14.7±0.4 | 15.9±0.5 | **0.7** | 19.0±0.4 | 16.7±0.5 | 15.5±0.4 | 14.6±0.4 | 16.4±0.4 | **0.9** |
| Cu-*K* | 35.2±0.8 | 34.3±0.8 | 37.2±0.9 | 34.9±0.7 | 35.4±0.8 | **1.3** | 32.7±0.6 | 30.0±0.7 | 33.1±0.7 | 33.2±0.6 | 32.2±0.7 | **1.5** |
| Sr-*K* | 23.7±1.9 | 25.1±3.1 | 23.8±3.3 | 25.3±1.7 | 24.5±2.5 | **0.8** | 24.1±2.4 | 25.7±2.0 | 26.6±1.7 | 24.4±2.3 | 25.2±2.1 | **2.2** |
| Zr-*K* | 4.1±2.0 | 7.2±2.2 | 3.2±2.4 | 4.9±1.9 | 4.8±2.1 | **1.7** | 5.4±1.7 | 5.6±2.1 | 4.1±1.9 | 6.8±1.6 | 5.5±1.8 | **1.6** |
| Ba-*L* | 15.3±0.6 | 13.5±0.6 | 14.4±0.6 | 15.0±0.5 | 14.5±0.6 | **0.8** | 15.1±0.5 | 16.8±0.6 | 17.0±0.5 | 17.5±0.5 | 16.6±0.5 | **0.9** |
| La-*L* | 3.7±0.3 | 3.2±0.6 | 3.7±0.4 | 2.9±0.5 | 3.4±0.4 | **0.4** | 2.8±0.3 | 3.6±0.4 | 2.7±0.5 | 2.4±0.5 | 2.9±0.4 | **0.6** |
| Pb-*L* | 1.0±0.3 | 1.3±0.3 | 1.2±0.4 | 1.3±0.3 | 1.2±0.3 | **0.2** | 1.0±0.3 | 1.6±0.3 | 1.1±0.3 | 1.3±0.3 | 1.2±0.3 | **0.3** |

**Table 2.** Quantitative content of elements in atomic % of various spots related to Fig. 7 (a,b).

| No# | Atomic % | | | | | | |
|---|---|---|---|---|---|---|---|
| | Ca-K | Cu-K | Sr-K | Zr-K | Ba-L | La-L | Pb-L |
| 1 | 10.5±0.6 | 56.9±1.0 | 26.5±1.2 | 0.9±0.7 | 4.6±0.3 | - | 0.3±0.2 |
| 2 | 29.0±1.1 | 0.2±0.3 | 11.5±1.2 | - | 59.3±1.3 | - | - |
| 3 | 7.3±0.7 | 29.5±1.2 | 19.3±1.8 | 0.2±1.6 | 43.1±1.4 | 0.6±1.3 | - |
| 4 | 3.2±0.5 | 14.1±0.7 | 6.8±0.8 | 62.3±2.1 | 11.0±1.0 | 0.1±0.5 | 2.3±0.4 |
| 5 | 5.6±1.3 | 40.9±1.6 | 10.8±1.8 | 0.5±1.6 | 3.2±3.7 | 36.2±2.7 | 0.1±0.7 |
| 6 | 7.2±0.8 | 49.0±1.3 | 13.3±1.0 | 1.2±0.9 | 14.1±1.4 | 14.2±1.4 | 1.0±0.3 |
| 1 | 2.2±0.3 | 53.9±1.3 | 43.2±1.9 | - | 0.5±0.3 | 0.2±0.3 | 0.1±0.3 |
| 2 | 2.3±0.5 | 58.1±1.4 | 35.6±2.0 | - | 4.0±0.5 | - | - |
| 3 | 1.9±0.4 | 58.8±1.3 | 34.9±1.9 | - | 3.9±0.5 | - | 0.5±0.4 |

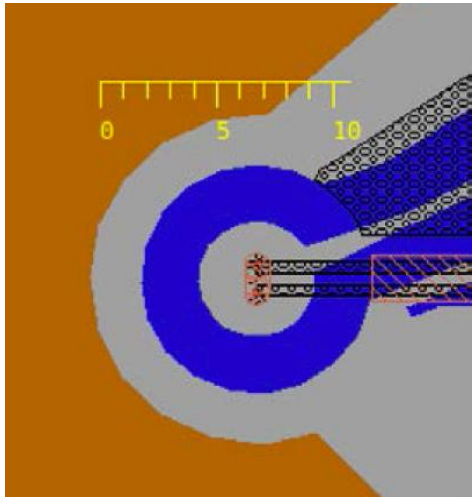

**Figure 1.** Layout of the scanning SQUID susceptometer with high spatial resolution used for identifying low concentrations of superconducting phases in inhomogeneous samples. The field coil is blue and the pickup loop leads are black. The cross-bar between the two leads is in red. The scale unit is 1 μm.

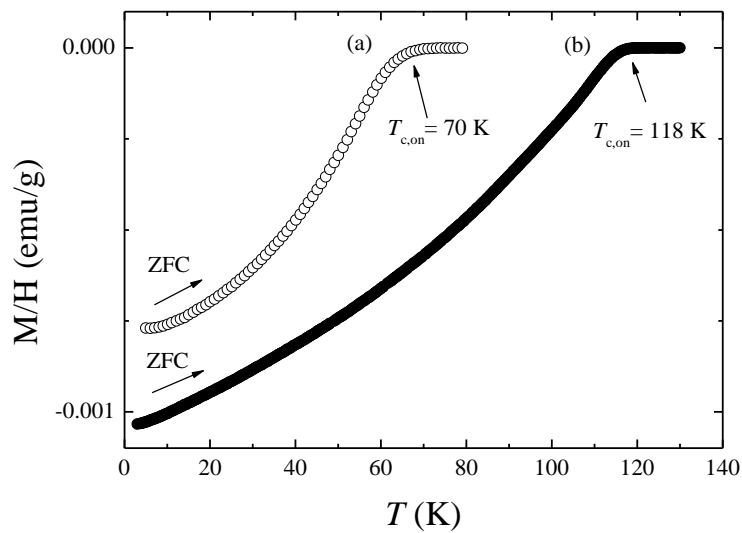

**Figure 2.** Temperature dependence of the magnetic susceptibility after zero-field cooling (ZFC) for a B-3 (a) and a B-10 (b) sample. Data obtained at an applied magnetic field of 50 Oe.

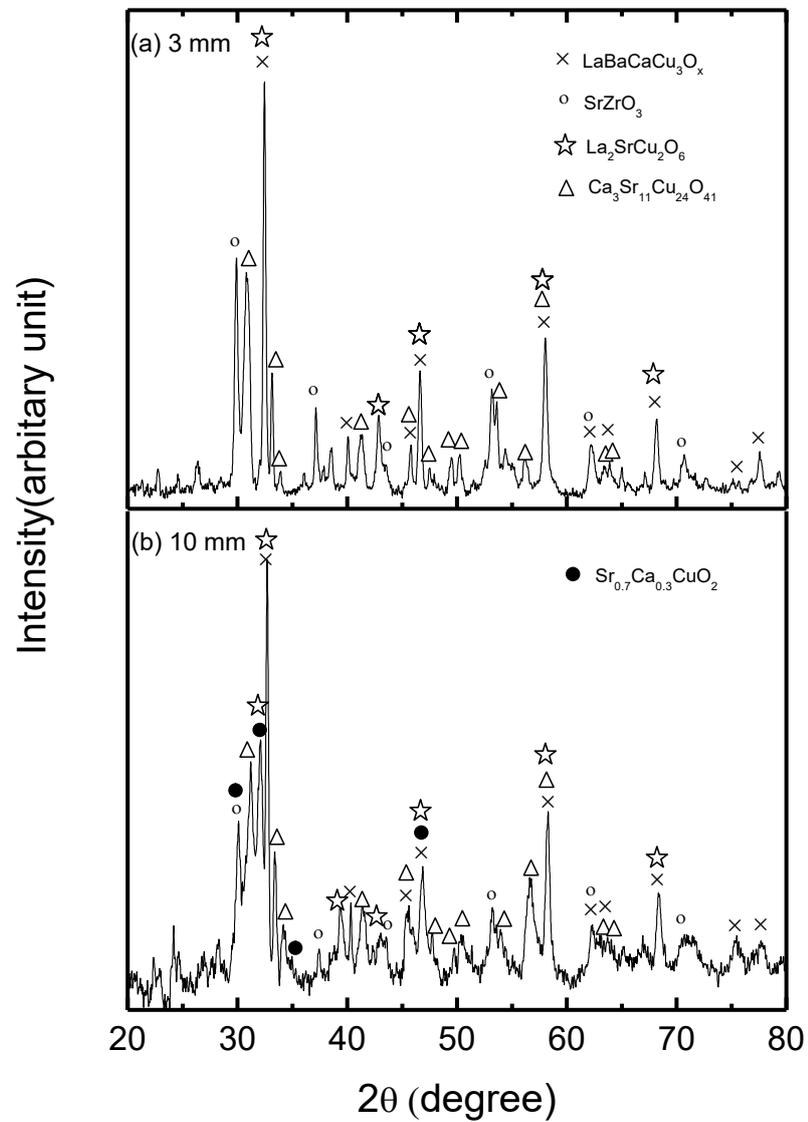

**Figure 3.** XRD pattern of samples prepared by (a) 3 mm ball sizes resulting in a low $T_c$, and (b) 10 mm ball sizes resulting in a high $T_c$.

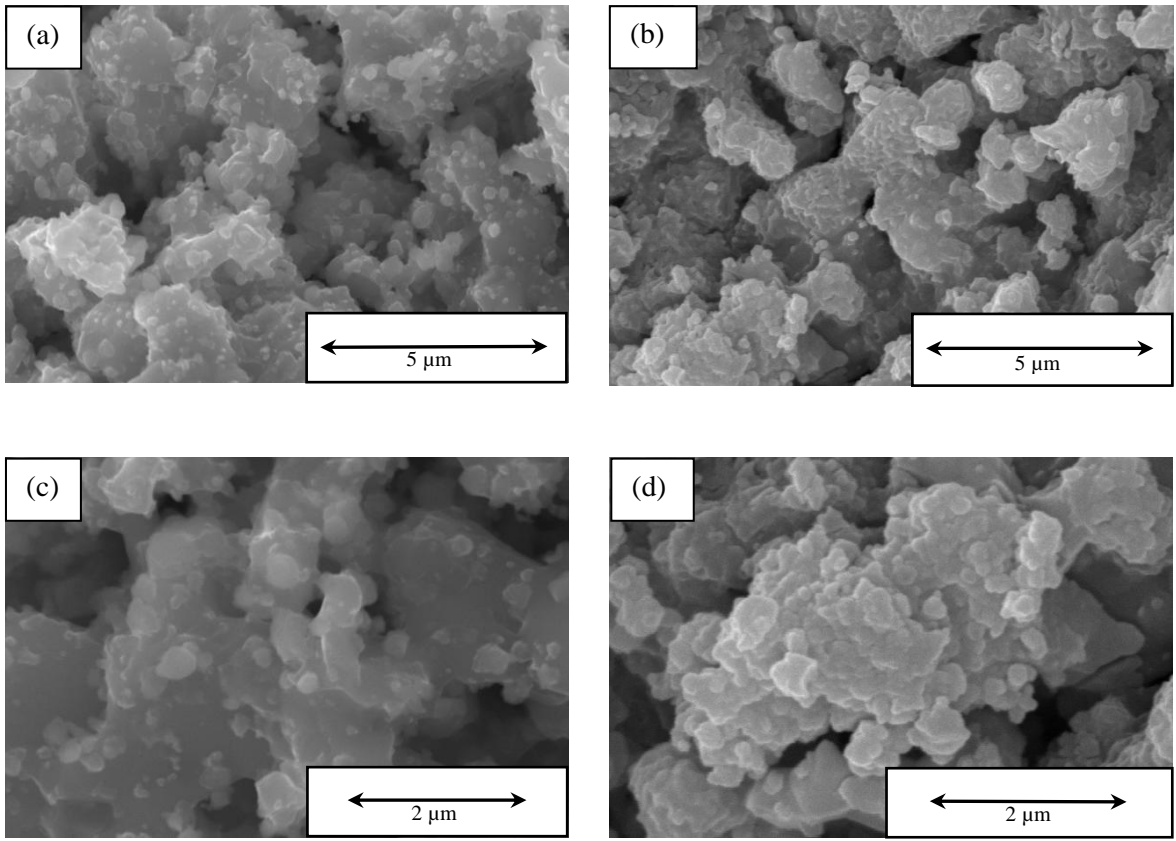

**Figure 4.** SEM morphology of samples prepared with different ball mill sizes: (a), (c) 3 mm and (b), (d) 10 mm.

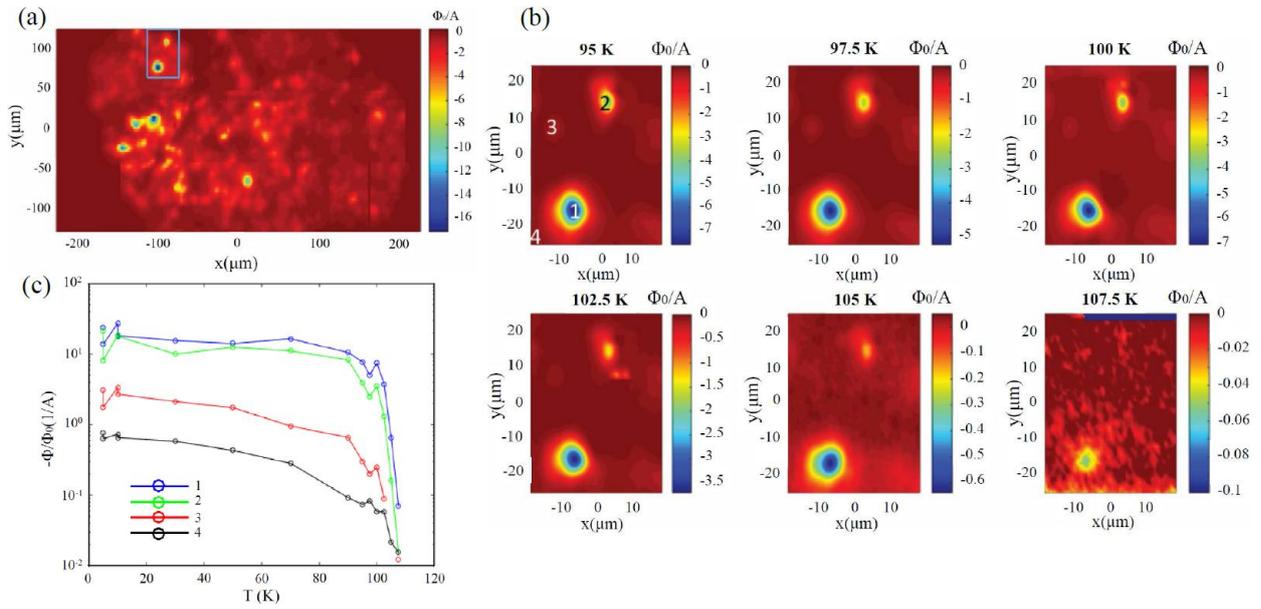

**Figure 5.** (a) Susceptibility image of a 300 × 200 μm$^2$ polished superconducting particle, imaged at 4.2 K. The area outlined by a square (including the region of the largest diamagnetic shielding in the sample) was imaged as a function of temperature (b). The diamagnetic susceptibility of the 4 regions (point 1 – blue, point 2 – green, point 3 – red and point 4 – black) labelled in the 95 K data (b) is plotted as a function of temperature in Figure 4 (c).

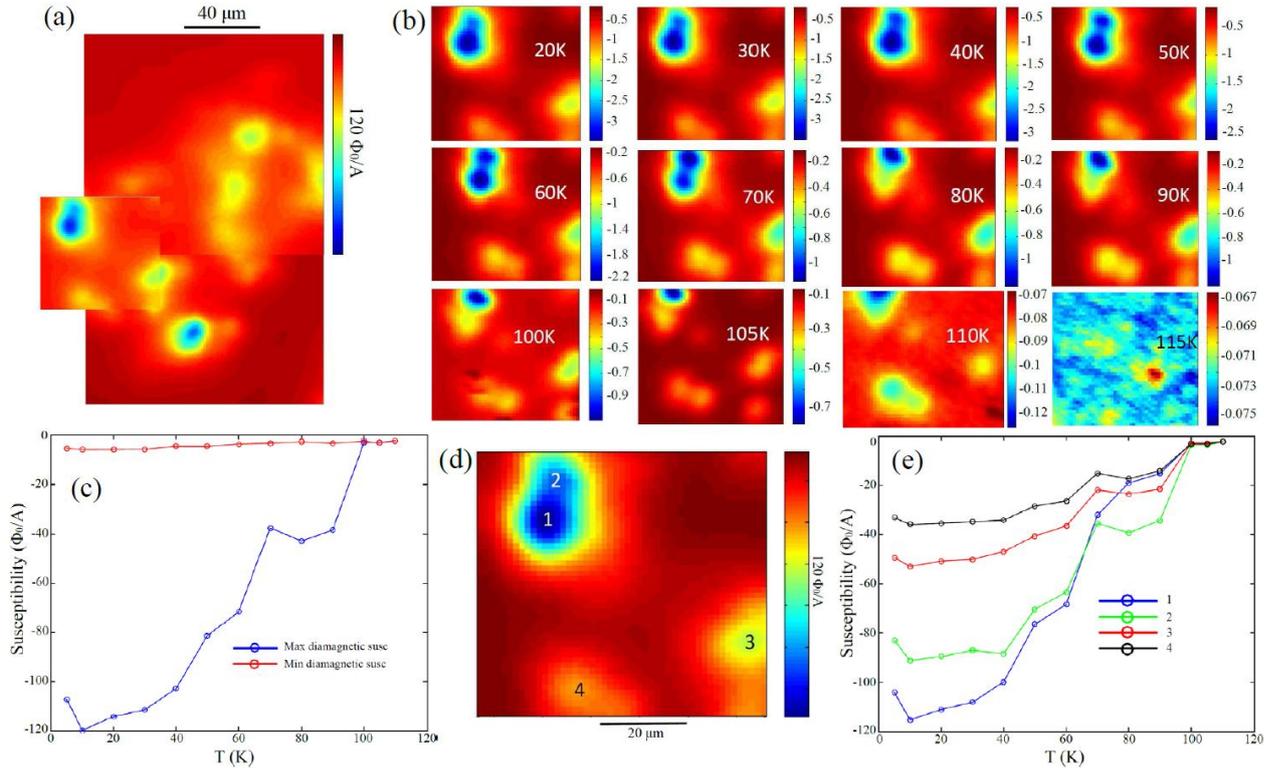

**Figure 6.** (a) Susceptibility mosaic image at 4.2 K of the same sample as in Fig. 5, but using a scanning SQUID susceptometer with a larger (5 micron radius) pickup loop size. The area chosen for a detailed study of the temperature dependence is shown to the left. (b) Susceptibility images of a 70 μm × 70 μm area of the sample at selected temperatures. (c) Plots of the smallest diamagnetic shielding (red) and largest diamagnetic shielding (blue) in the images of (b) as a function of temperature. (e) The temperature dependences of the local maxima in diamagnetic susceptibility in the locations labeled in (d). According to EDS analysis these superconducting local regions can be assigned to the "$(Sr,Ca,Ba)_{0.7-1.0}CuO_2$" and "Pb-1223" phases.

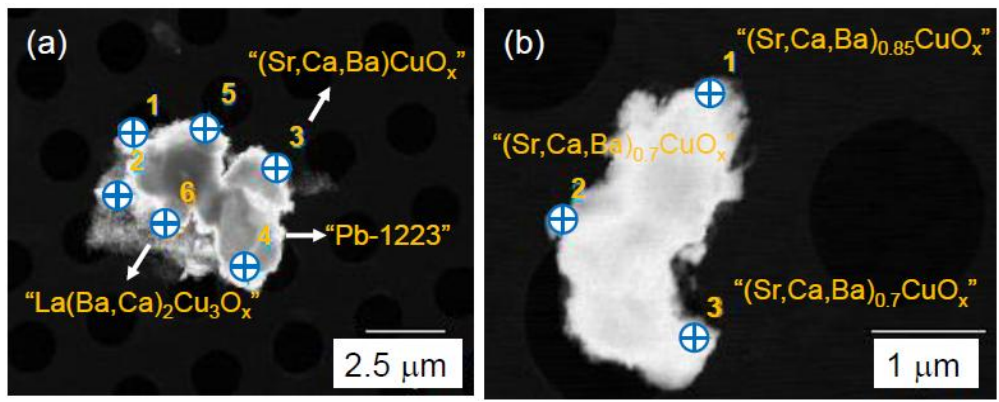

**Figure 7(a,b).** TEM images of a captured superconducting particles with 6 and 3 spots in which EDS analysis was done. Five of them can be ascribed as ''$(Sr,Ca,Ba)_{0.7-1.0}CuO_x$'' (#3 in (a) and #1, #2, #3 in (b)), ''Pb-1223'' (#4) and ''$La(Ba,Ca)_2Cu_3O_x$'' (#6) phases.